\documentclass[11pt,a4paper]{article}
\usepackage[nohead, nomarginpar, margin=1in, foot=.25in]{geometry}

\usepackage{graphicx}
\usepackage{color}

\usepackage{amsmath}
\usepackage{amsmath,bbm}

\usepackage{amssymb,bm}

\usepackage{csquotes}
\usepackage{subcaption}

\usepackage[T1]{fontenc}
\usepackage[utf8]{inputenc}
\usepackage{authblk}

\usepackage{soul}
\usepackage[LGRgreek]{mathastext}

\usepackage{url}

\usepackage{lineno}

\DeclareSymbolFont{greekletters}{OML}{cmr}{m}{it}
\DeclareMathSymbol{\varrho}{\mathalpha}{greekletters}{"25}

\usepackage{cite}
\date{}

\def\@listcomma@comma{\@ifnum{\@tempcnta>\tw@}{,}{}}

\begin{document}


\title{Calculating charged particle observables using modified Wood Saxon model in HIJING for U+U collisions at $\sqrt{s_{NN}}$ = 193 GeV}
\author[1,2]{S. K. Tripathy}
\author[3, \footnote{younus.presi@gmail.com}]{M. Younus}
\author[2]{Z. Naik}
\author[1]{P. K. Sahu}

\affil[1]{Institute of Physics, HBNI, Sachivalaya Marg, Bhubaneswar 751005, India}
\affil[2] {Sambalpur University, Jyoti Vihar, Burla, Sambalpur 768019, India }
\affil[3] {Indian Institute of Technology Indore, Simrol, Indore 453552, India}

\maketitle

\begin{abstract}
We have implemented spherical harmonics in default Wood Saxon distribution of the HIJING model and calculated various physical observables such as transverse momentum, charged particle multiplicity, nuclear modification factor and particle ratios for charged particles at top RHIC energy with collisions of Uranium (U) nuclei. Results have been compared with available experimental data. We observe that, a particular type of collision configuration can produce significant magnitude change in observables. We have noticed that the tip-tip configuration shows higher magnitude of particle yield in central collisions, while the body-body configuration shows higher value in the cases of peripheral collisions, with the flip in the trend occurring for the mid-central U+U collisions. We observe that particle ratios are independent of configuration type.
\end{abstract}


\noindent
\textit{Keywords : Monte Carlo simulations, deformed nuclei, Charged particle production} \\
\textit{PACS: 21.60.Ka, 25.75.Dw} \\


\section{Introduction}
\label{intro}
Experiments with relativistic heavy ion collisions have shown formation of a hot and dense system of deconfined quarks and gluons commonly known as quark gluon plasma (QGP). Recent experiments with state-of-the-art technology involving gold (Au) and lead (Pb) nuclei are being conducted both at RHIC-BNL and LHC-CERN.  While experimental data tends to reconstruct the entire heavy ion collision scenario from finally produced hadrons and leptons etc., theoretical and phenomenological models could calculate the final outcome by incorporating various analytical methods starting from the initial conditions of relativistic heavy ion collisions. It has been observed that initial conditions determine the evolution of quark gluon plasma and many of the final observables such as particle flow and correlations bear the signatures of such effects \cite{initial_condn_1,initial_condn_2}. Thus precise determination of initial conditions should play a vital role in explaining experimental observables. Theoretically it is assumed that heavy ions such as Au or Pb are almost spherical and have negligible or zero deformations. In order to have precise determination of the nature of initial conditions for different configurations, intrinsically deformed nuclei such as uranium (U) plays a vital role. The initial cold nuclear matter effects such as partons or nucleons multi-scattering when two nuclei collide may show certain orientation dependencies. Similarly, initially produced particles should also depend upon number of binary collisions of nucleons, and that too should depend on the orientations of the two colliding nuclei.

In case of Au (mildly deformed) or Pb (zero deformation), we use Wood-Saxon (WS) distribution to define the distribution of nucleons within the nucleus. For the intrinsically deformed nucleus such as prolate shaped uranium, U, we have included $n^{th}$ ordered deformation parameter, $\beta_n$, associated with spherical harmonics, $Y_{nl}(\theta)$, in the WS function~\cite{mws_1,mws_2, mws_3,mws_4,mws_5}. The modified distribution is called modified Wood-Saxon (MWS) density distribution for the nucleons within the nucleus. MWS distribution is used as initial nuclear matter density in U+U collision in order to explain the experimental data. We have tried to use MWS in the context of Glauber formalism within the HIJING model to calculate charged particle multiplicities, transverse momentum distribution of charged particles etc., for tip on tip, body on body and random configuration collisions of uranium nuclei.

The paper is organised as follows. In Sec.\,\ref{sec:mws_calc} we have shown optimisation of MWS parameters to get the best values for the charged particles distribution, using HIJING code. This is followed by Sec.\,\ref{sec:results_disc} on the results and their discussion. We finally summarise our results on Sec.\,\ref{sec:conclusion}.

\section{Formalisms}
\label{sec:mws_calc}

\subsection*{Modified Wood Saxon distribution (MWS) distribution}

In phenomenological approach for many body system, nuclear charge density is usually interpreted in a three parametric Fermi distribution \cite{density_fermitype}. 
\begin{equation}
\rho (r) = \rho_0 [\frac{1+w(r/R)^{2}}{1+exp[(r-R)/a]}]\,.\\
\label{eqn1}
\end{equation}
Here $\rho_0$ is the nuclear matter density in the centre of the nucleus, R is  the radius of the nucleus from its centre, and it is assumed that nuclear matter density reduces to the half of its maximum value at this distance. The parameter, a, is the skin depth or surface thickness, r is a position parameter and distance of any point from centre of the nucleus, 
and w is the deviation from a smooth spherical surface.

$\rm{Au}^{197}$ or $Pb^{208}$ nucleus is assumed here to have uniform distribution of nucleons in its approximately spherical volume so that $\displaystyle w$ can be taken to be zero.
This reduces~eqn.\ref{eqn1} to the popular Wood-Saxon \cite{wdsx} distribution, which is used in most of the event generators for heavy-ion collisions. This may be written as:\\
\begin{equation}
\rho (r) = \frac{\rho_0}{1+exp[(r-R)/a]}\,.\\
\label{eqn2}
\end{equation}

When we use an axially symmetric or prolate deformed nucleus (viz. $U^{238}$), nuclear radius may be modified to include spherical harmonics as well. The deformed Wood-Saxon nuclear radius \cite{wdsx_deform} may be written as:
\begin{equation}
R_{A\Theta} = R[1+ \beta_2 Y_{20}(\theta)+\beta_4 Y_{40}(\theta) ]
\label{eqn3}
\end{equation}
Where the symbols $\beta_i$ are deformation parameters. We have used deformation parameters $\beta_2$ = 0.28 and $\beta_4$ =0.093 \cite{mws_2, beta,atomic_data_table} in our calculations for uranium nuclei. 

The spherical harmonics, $Y_{20}$, is given by~\cite{speherical_harmonic}, \\
$Y_{20}(\theta) = \frac{1}{4} \sqrt{\frac{5}{\pi}}(3 \ cos^2\theta -1)$ and $Y_{40}(\theta) = \frac{3}{16\sqrt\pi}(35 \ cos^4\theta -30 \ cos^2\theta+3)$\,.\\

As the density distributions cannot be used to assign individual nucleon positions \cite{thesis_chris}, one must have to construct probability density function from it by sampling with differential volume element $r^2 sin \theta \ dr \ d\theta \ d\phi$ \cite{ptribedy}. For random or unpolarized orientation of nuclei, one has to take care of both polar angle (angle between major axis and beam axis, $\Theta \in [0, \pi)$) and azimuthal angle (angle between major axis and impact parameter, $\Phi \in [0, 2\pi)$). The first one sampled according to $sin\Theta$ probability distribution and later one with uniform distribution. Both target and projectile nuclei are rotated event by event in azimuth and polar space. Impact parameter sampling has been done linearly.
In this paper, random orientation means, unpolarized and averaged value over random $\Theta$ and $\Phi$. Tip orientation is for $\Theta =0$ and $\Phi \in [0, 2\pi)$) while Body orientation means  $\Theta = \pi/2$ and $\Phi$ =0 \cite{define_config}.

\subsection*{Event generator model, HIJING}
Heavy Ion Jet INteraction Generator (HIJING) \cite{hijing} program is designed explore possible initial conditions that may occur in relativistic heavy-ion collisions. HIJING assumes nucleus-nucleus collision which can be decomposed into binary nucleon-nucleon collisions. It  uses a three-parameter Wood-Saxon nuclear matter density to compute the number of binary collisions at a given impact parameter. Between each pair of colliding nucleons, impact parameter is calculated using their transverse positions. Eikonal formalism, which uses straight line trajectories between two nucleons, is used to calculate probability of collision. 
Once all binary collisions are processed, then scattered partons in the associated nucleons are connected with the corresponding valence quarks to form string systems. Finally, particles are formed from the fragmentation of these strings. PYTHIA subroutine is used to generate kinematic variables for each semi/hard scattering.

We have implemented MWS formalism in HIJING model as mentioned in Eq. \ref{eqn3}. 
We have taken top RHIC energy (U+U $\sqrt{S_{NN}}$=193 GeV) and charged particles in our calculations. Impact parameter has been used to calculate centrality of collision.

Participating and colliding nucleons numbers, $N_{part}$, and $N_{coll}$ have been calculated using Glauber model in optical approximation. One can show from the Ref \cite{miller}, these values as function of impact parameter or centralities,:

\begin{equation}
T_{AB}(b)=\int{T_A({s})\,.T_B({|\bf{s}-\bf{b}|})\,d^2s}\,,
\label{TAB}
\end{equation}
where, $T_{AB}(b)$ is the nucleus overlap function at a given impact parameter, $b$. Therefore, the no. of nucleon binary collisions in $A+B$ collisions, is given by

\begin{equation}
N_{coll}(b)=A.B.T_{AB}(b).\sigma_{NN}\,,
\label{ncoll}
\end{equation}

and the no. of participants is given by,

\begin{eqnarray}
N_{part}(b)&=&A.\int{T_A({s}).\big\{1-[1-T_B({|\bf{s}-\bf{b}|})\sigma_{NN}]^{B}\big\}}.d^2s\nonumber\\
&+&B.\int{T_B({|\bf{s}-\bf{b}|}).\big\{1-[1-T_A({s})\sigma_{NN}]^{A}\big\}}.d^2s\,,
\end{eqnarray}

where $\sigma_{NN}$ is the inelastic nucleon-nucleon cross section measured for a given collision energy of two nucleons. A and B are the mass numbers of the two colliding nuclei.

\section{Results and discussions}
\label{sec:results_disc}

\begin{figure}
\centering
\includegraphics[scale=0.5]{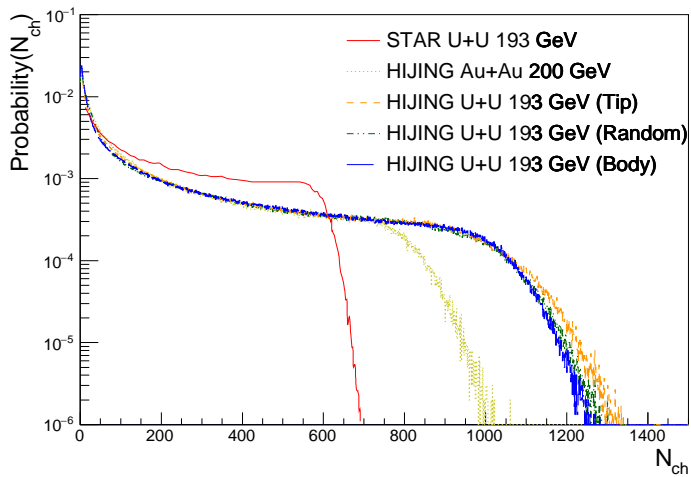}
\caption{(Color online) Charged particle distribution for U+U collisions at $\sqrt{s_{NN}}$ = 193 GeV and Au+Au collisions at $\sqrt{s_{NN}}$ = 200 GeV from HIJING along with experimental data for U+U at $\sqrt{s_{NN}}$ = 193 GeV \cite{star_nch}. Two specific orientations viz. body-body and tip-tip are shown along with random orientation.}
\label{fig:compare_nch} 
\end{figure}

In Fig \ref{fig:compare_nch}, we have shown charged particle multiplicity ($N_{ch}$) distribution from HIJING for U+U and Au+Au collisions. Only minimum-bias and mid-rapidity ($|\eta| < 0.5$) particles are considered here. Although $N_{ch}$ shapes are consistent in all formalism, MWS random distribution lies between, body-body and tip-tip configurations. While body-body configuration yields least no. of charged particles, $n_{ch}\,\approx$ 1250 among all configurations, tip-tip configuration gives $\sim$10\% higher charged particles than body-body configuration and gives the maximum value. We observe that MWS distribution with random orientation of nuclei, gives slightly more results than body-body. 
In the same figure, we have plotted the charged particle distribution for Au+Au collisions at $\sqrt{s_{NN}}$ = 200 GeV and compared it with U+U collisions. The magnitude of the $N_{ch}$ at the high multiplicity region in case of Au+Au collisions is less than all of the orientations of U+U collisions . We have also compared our results with STAR experiments's preliminary data \cite{star_nch}. It shows highest multiplicity of $N_{ch} \ \sim$ 700.

\begin{figure}
\centering
\begin{subfigure}{.5\textwidth}
  \centering
  \includegraphics[width=\linewidth]{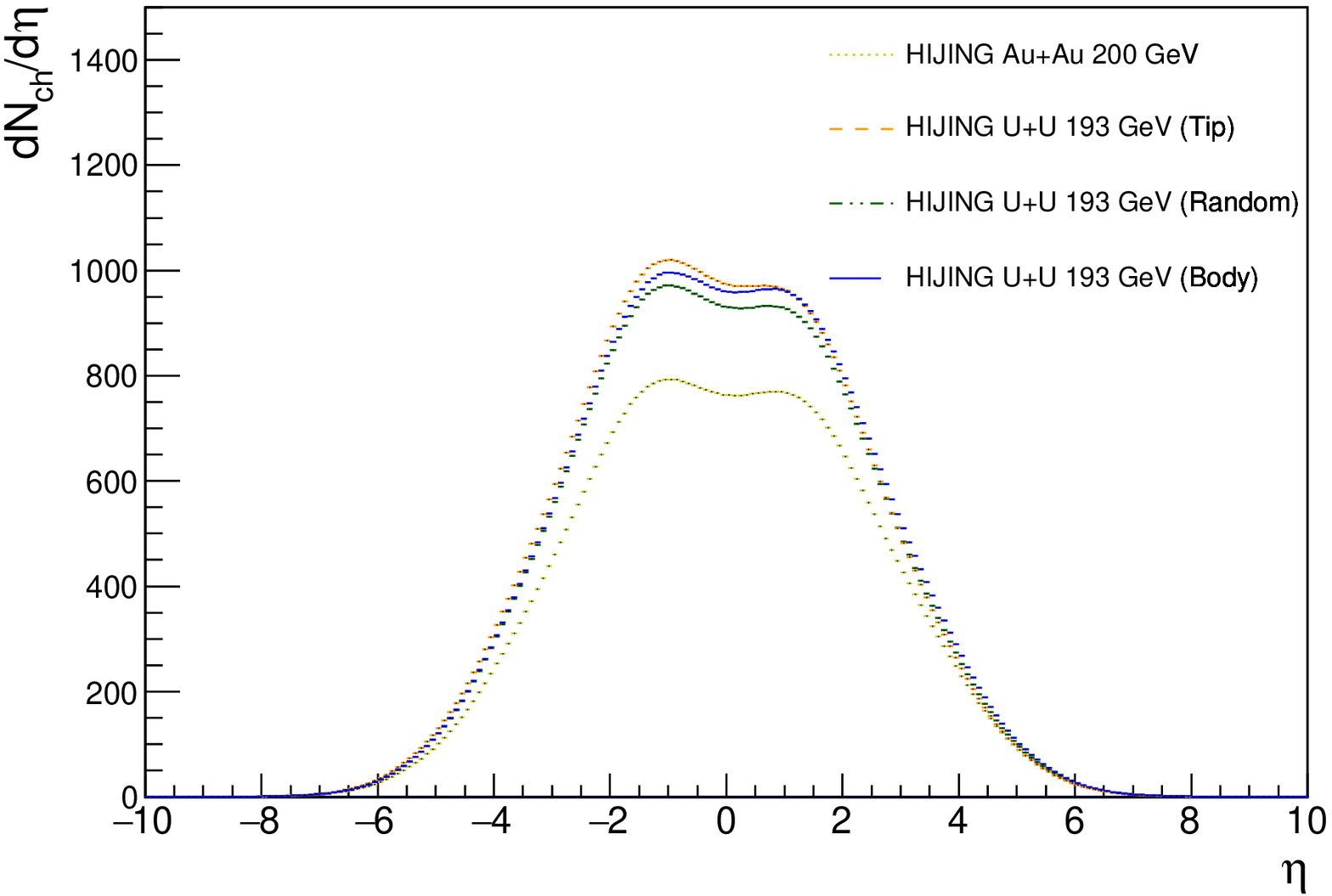}
  \caption{}
  \label{fig:compare_eta}
\end{subfigure}%
\begin{subfigure}{.5\textwidth}
  \centering
  \includegraphics[width=\linewidth]{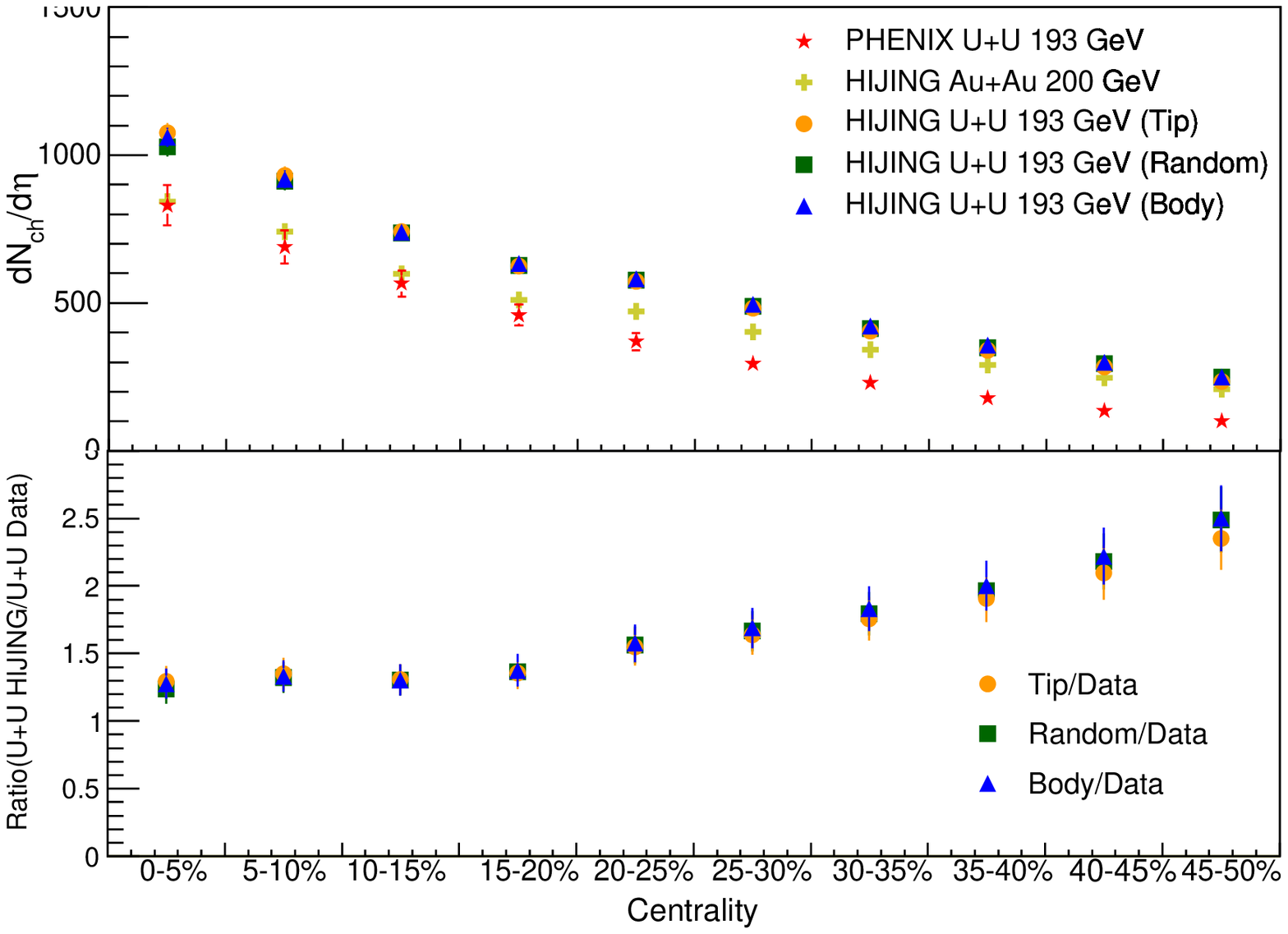}
  \caption{}
  \label{fig:compare_dnchdeta}
\end{subfigure}
\caption{(Color online) In Fig \ref{fig:compare_eta} Pseudo-rapidity distribution from HIJING is shown for U+U at $\sqrt{s_{NN}}$ = 193 GeV and Au+Au at $\sqrt{s_{NN}}$ = 200 GeV for 0-5\% centrality. Two specific orientations viz. body-body and tip-tip are shown along with random orientation. Fig \ref{fig:compare_dnchdeta} shows $dN_{ch}/d\eta$ distribution from HIJING along with experimental data  for U+U collisions at $\sqrt{s_{NN}}$ = 193 GeV \cite{phenix_dnchdeta} as a function of centrality. Au+Au 200 GeV results are shown along with this. In the bottom plot, ratio of HIJING to data in U+U 193 GeV are shown.}
\label{fig:compare_eta_dnchdeta}
\end{figure}

In Fig \ref{fig:compare_eta}, we have shown pseudo-rapidity distribution ($dN_{ch}/d\eta$) in U+U and Au+Au most central collisions (0-5\%) from HIJING.
In Fig \ref{fig:compare_dnchdeta}, we have $dN_{ch}/d\eta$ as a function of centrality. We observe that most central collision produces $dN_{ch}/d\eta$ $\sim$ 1100, while most peripheral collision produce $\sim$ 300 for U+U collisions. Tip-tip collision produces highest number of particles in most central collisions, while for the peripheral collisions, body-body configuration produces slightly more number of particles than the tip-tip. This trend however reverses around 10-20\% centrality. Here, too we have shown pseudo-rapidity distribution of charged particles from Au+Au collisions as a comparison. 
Ratios of HIJING results to experimental published data  \cite{phenix_dnchdeta} show increasing trend at higher centralities.

\begin{figure}
\centering
\includegraphics[scale=0.5]{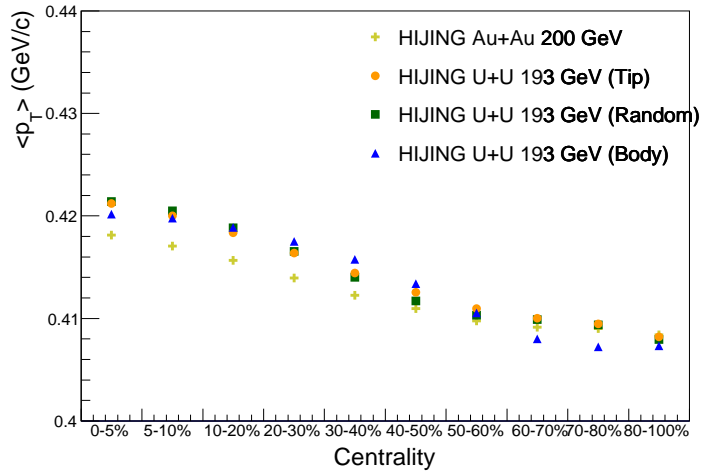}
\caption{(Color online) $<p_{T}>$ distribution from HIJING for U+U collisions at $\sqrt{s_{NN}}$ = 193 GeV and Au+Au collisions at $\sqrt{s_{NN}}$ = 200 GeV. Two specific orientations viz. body-body and tip-tip are shown along with random orientation.}
\label{fig:compare_meanpt} 
\end{figure}

In Fig \ref{fig:compare_meanpt}, we have plotted $<p_{T}>$ as a function of centrality. 
We see drop in $<p_{T}>$ from central to peripheral collision. Together with this, we have almost 5\%  difference in magnitudes between various orientations. However, the trend of the distributions shows a qualitative similar behaviour.In earlier paper by Rihan et al~\cite{mws_2}, where in the AMPT formalism, the dependence of average transverse momentum on orientation becomes more visible for more central collisions. We will keep investigating this difference from our work. In our current work, $<p_{T}>$  is 0.42 GeV for most central collisions, while it was 0.41 GeV for most peripheral collisions. Similar to Fig \ref{fig:compare_dnchdeta},  $<p_{T}>$  from tip-tip collision found to be higher for most central collision.  Here mean $p_T$ of the charged particles observed from Au+Au collisions is similar in magnitude to the tip-tip configurations for peripheral U+U collisions. 

\begin{figure}
\centering
\includegraphics[scale=0.6]{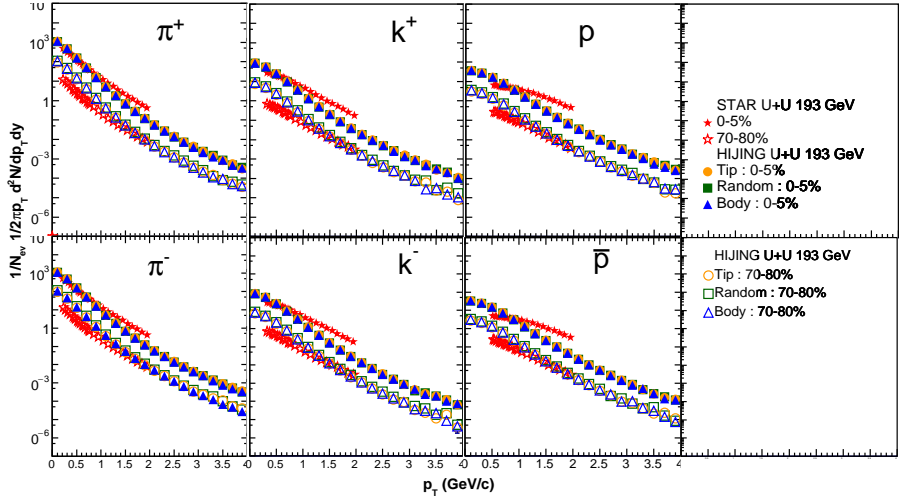}
\caption{(Color online) Transverse momentum distribution from HIJING and preliminary experimental data  \cite{star_spectra} for U+U collisions at $\sqrt{s_{NN}}$ = 193 GeV. Two specific orientations viz. body-body and tip-tip are shown along with random orientation.}
\label{fig:compare_ptspectra} 
\end{figure}

In Fig \ref{fig:compare_ptspectra} we have plotted transverse momentum spectra at the most central (0-5\%) and most peripheral collisions (70-80\%) using MWS distribution for charged particles produced in U+U collisions and compared with recent experimental preliminary result \cite{star_spectra}. Experimental data is is up to $p_{T}$ < 2 GeV. HIJING results for pions are in trend with experimental result, while protons results have the most mismatch with the data. The mismatch of our results with experimental may be due to the absence of secondary multiple interactions of partons in HIJING. We are currently looking into the possible reasons behind this and will report in future.

\begin{figure}
\centering
\includegraphics[scale=0.5]{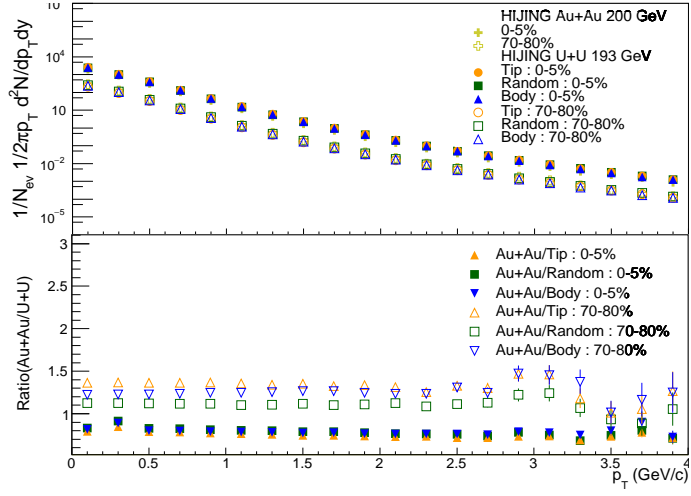}
\caption{(Color online) Transverse momentum spectra from HIJING for U+U collisions at $\sqrt{s_{NN}}$ = 193 GeV and Au+Au collisions at $\sqrt{s_{NN}}$ = 200 GeV. Two specific orientations viz. body-body and tip-tip are shown along with random orientation.}
\label{fig:compare_spectra_auau} 
\end{figure}

In Fig \ref{fig:compare_spectra_auau} we have plotted transverse momentum spectra for most central (0-5\%) and most peripheral collisions (70-80\%) in U+U collisions at $\sqrt{s_{NN}}$ = 193 GeV and Au+Au collisions at $\sqrt{s_{NN}}$ = 200 GeV at top plot. In the bottom, ratios of these various types of configuration to the Au+Au results are presented. We observe that, central collisions in U+U yield higher magnitude, while Au+Au results give higher magnitude for peripheral collisions. Ratios of different configurations of U+U with Au+Au for central collisions give almost similar values (less than unity), while peripheral collision shows a configuration dependent magnitude; tip-tip configuration being highest among all.  Although there is fluctuation in the ratio with Au+Au, we observe the ratio goes smoothly with transverse momentum and centrality.
The flip in magnitude of the ratio, might be reflected in physics observables, which use both centrality values, viz. as nuclear modification factor, which we now  discuss in the following paragraph.

\begin{figure}
\centering
\includegraphics[scale=0.5]{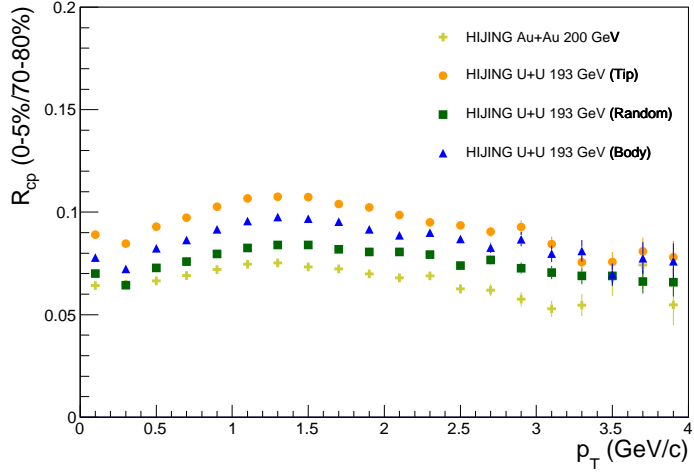}
\caption{(Color online) Nuclear modification factor distribution from HIJING for U+U collisions at $\sqrt{s_{NN}}$ = 193 GeV and Au+Au collisions at $\sqrt{s_{NN}}$ = 200 GeV. Two specific orientations viz. body-body and tip-tip are shown along with random orientation.}
\label{fig:compare_rcp} 
\end{figure}

In Fig \ref{fig:compare_rcp} we have presented nuclear modification factor as a function of transverse momentum using MWS distribution. We have used the following definition of nuclear modification ($R_{CP}$) in the current calculations, where peripheral collisions is assumed to be devoid of any thermalized systems,
\begin{equation}
R_{CP} = \frac{d^{2}N/dp_{T}dy/<N_{coll}^{cent}>}{d^{2}N/dp_{T}dy/<N_{coll}^{Perph}>}\,.
\end{equation}
Where $<N_{coll}^{cent}>$ and $<N_{coll}^{Perph}>$ are average number of binary collisions in central (0-5\%) and  peripheral (70-80\%) collisions, which have been calculated using the same mathematical approach of Optical Glauber Model used in HIJING.

We have used the nuclear shadowing parametrization incorporated in HIJING as well as energy loss mechanisms defined within the code. The energy loss depends on the rate of induced bremsstrahlung when any particle moves through the hot and dense partonic matter. The interaction points with the medium particles to calculate energy loss $dE/dl$, are determined by the probability,
\begin{equation}
dP=\frac{dl}{\lambda_s}e^{-dl/\lambda_s}\,,
\end{equation}
where $\lambda_s$ is the mean free path and $dl$ is the elementary path-length traveled by the particle in QGP. The energy loss is then given by, $\displaystyle \Delta E=dl*dE/dl$.

Interestingly we observe, all configuration types of U+U collision system show modification to lesser magnitude, than Au+Au collision system. This might be attributed to the fact discussed from Fig \ref{fig:compare_spectra_auau}, where we have shown that $p_{T}$ spectra in $Au+Au ^{peripheral} \ > \ U+U^{peripheral}$, while the reverse is true for central collisions.

The fact we observed from Fig~\ref{fig:compare_rcp} is that the energy loss depends on the orientation of the colliding uranium nuclei, where body-body configuration gives more suppression than tip-tip. On the other hand the cold nuclear matter effects such as shadowing which manifests itself at low $p_T$ (< 1 GeV), has affected the particle production at very low $p_{T}$ as usual. However similarity in the shapes of the distribution shows negligible effects of various orientations on the shadowing \cite{Sarsour:2016rzm}.

 \begin{figure}
\centering
\includegraphics[scale=0.5]{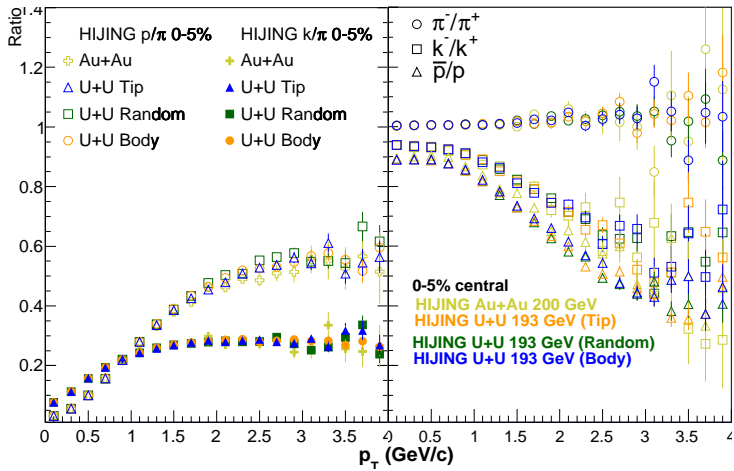}
\caption{(Color online) Particle ratios from HIJING for U+U at $\sqrt{s_{NN}}$ = 193 GeV and Au+Au at $\sqrt{s_{NN}}$ = 200 GeV. Two specific orientations viz. body-body and tip-tip are shown along with random orientation.}
\label{fig:compare_ratio} 
\end{figure}

In Fig \ref{fig:compare_ratio} we have plotted particle ratios as function of $p_{T}$. In the left plot, $p/\pi$ and $k/\pi$ ratio for most central collisions (0-5\%) as a function of $p_{T}$. From $p_{T}$ > 1 GeV onward,  the trend of $k/\pi$ ratio reverses and goes towards saturation faster than $p/\pi$ ratio. In right plot, we have presented anti-particle to particle ratios $\pi^{-}/\pi^{+}$, $k^{-}/k^{+}$ and $\bar{p}/p$ as a function of $p_{T}$ in most central collisions (0-5\%). While pions give an almost flat ratio close to unity, others particle ratios decrease from unity for $p_{T}$ > 1 GeV.  We do not observe any orientation or collision configuration dependencies in either of the ratio plots. 
Similarly, when the particle ratios from Au+Au collisions is compared to U+U collisions at almost same c.m. energies, we do not find any discernible effects of the system (QGP) on the particle ratios. We will continue to investigate the effects of system sizes on the particle production in our future works.

\section{Summary}
\label{sec:conclusion}
In this paper we have implemented MWS distribution within HIJING formalism and calculated physical observables (viz. particle spectra and ratios) for charged particle at RHIC energy, U+U at $\sqrt{S_{NN}}$ = 193 GeV. 
We have used two particular orientations of colliding uranium nuclei, viz.  tip-tip and body-body along with random rotation. For comparison purposes, we have also shown results from Au+Au  $\sqrt{S_{NN}}$ = 200 GeV collisions. Comparisons are also made with experimental data wherever available. 

$N_{ch}$ distribution in random configuration gives closer values to that body-body configuration, however larger statistics might be required to discern the fine disagreements between the two. 
We also notice that body-body configuration yields lesser particles while tip-tip shows higher production of charged particles.  For central collisions, $dN_{ch}/d\eta$ in all configurations, gives similar deviation from data. While in peripheral collision, up to 30\% variation is observed from experimental data to HIJING results. $<p_T>$ shows similar trend among different configurations and also depends on centrality. 
$p_T$ spectra from HIJING shows similar trend for pion experimental data, while differ considerably for protons. 
We have found $R_{CP}$ depends upon the orientation and thus may imply that the extent of formation of hot and dense system depends upon the orientation itself. 
Particle ratios are found to be independent of collision configuration in most central collision. Furthermore, our studies with Au+Au as well as U+U collisions at the almost similar c.m. collision energies couldn't reveal any system size dependency on the observables such as particle ratios.  
However, energy loss suppression factor affecting $R_{CP}$ and transverse momentum spectra, may reveal this dependency to some extent.

Our study shows that, tip-tip configuration can give us higher number of produced particles than other configurations. 
It is also interesting to see that Au+Au systems give rise to more particles production than U+U systems for the peripheral collisions. \\
We aim to further extend this study in finding methods in selecting a particular type of configuration among all and give us indirect view of event-by-event analysis of U+U experimental data.

\section*{Acknowledgement}
Authors greatly thank Prithwish Tribedy, Physics Department, Brookhaven National Laboratory for his valuable discussion and suggestions on technical aspects of this paper.

\end{document}